\documentclass[11pt, oneside]{article}   	
\usepackage{geometry}                		
\usepackage{graphicx}				
\usepackage{amssymb}

\title{New non-linearity parameters of Boolean functions}
\author{Igor Semaev\\ Department of Informatics, University of Bergen}

\begin{document}
\maketitle

\begin{abstract}
The study of non-linearity (linearity) of Boolean function was initiated by Rothaus in 1976. The classical non-linearity of a Boolean function is the minimum Hamming distance of its truth table to that  of affine functions. 
In this note we introduce new "multidimensional"  non-linearity parameters $(N_f,H_f)$ for conventional and vectorial Boolean functions $f$ with $m$ coordinates in $n$ variables.
 The classical non-linearity may be treated as a 1-dimensional parameter in the new definition.  $r$-dimensional parameters for $r\geq 2$ are relevant to possible multidimensional extensions of the Fast Correlation Attack in stream ciphers and Linear Cryptanalysis in block ciphers. Besides we introduce a notion of  optimal vectorial Boolean functions relevant to the new parameters. For $r=1$ and even $n\geq 2m$ optimal Boolean functions  are exactly  perfect nonlinear functions (generalizations of Rothaus' bent functions) defined by Nyberg in 1991. By a computer search we find that this property  holds for $r=2, m=1, n=4$ too. That is an open problem for larger $n,m$ and $r\geq 2$. The definitions may be easily  extended to $q$-ary functions.       
\end{abstract}
\section{Conventional Boolean Functions}\label{Boolean}
Let $f(x)=f(x_1,\ldots,x_n)$ be a Boolean function (takes $0,1$-values) in $n$ Boolean variables $x=(x_1,\ldots,x_n)$. Let   $v$ denote its weight (the number of values $1$ in the truth table). One constructs a probability distribution $p$ on binary $n$-strings, such that $p_x=1/v$ if $f(x)=1$ and $p_x=0$ otherwise.

Let $r$ be a fixed number $1\leq r\leq n$ and  $U$  an $r\times n$ binary matrix of rank $r$. The matrix defines a linear transform from the space of $n$-bit strings to the space of $r$-bit strings. That induces  
  a probability distribution $q$ on binary $r$-strings. 
 Namely, $q_y=\sum_{y=Ux}p_x$, where the sum is over  $x$ such that $y=Ux$.  The distribution $q$ depends on $U$. For $r=n$ the distribution $q$ is a permuted distribution $p$ with a linear permutation defined by $U$.
 
  For $r=1$ the distributions (in a slightly different form) are used in  Correlation and Fast Correlation Attacks in stream ciphers, see \cite{MS}.
  The efficiency of  Correlation  
Attacks,  e.g., for a Filter Generator with a filtering function $f$, depends on the probability
$\mathbf{Pr}(Ux=f(x)).$ By the definition of $q=(q_0,q_1)$, one gets $\mathbf{Pr}(Ux=1,f(x)=1)=vq_1/2^n$ and $\mathbf{Pr}(Ux=0,f(x)=0)=1/2-vq_0/2^n$. 
So 
\begin{eqnarray}
\mathbf{Pr}(Ux=f(x))&=&\mathbf{Pr}(Ux=1,f(x)=1)+\mathbf{Pr}(Ux=0,f(x)=0)=\frac{1}{2}+\frac{v(q_1-q_0)}{2^n}.\nonumber\end{eqnarray}
 For $r\geq 2$ the distribution $q$ may potentially    
    be used in multidimensional extensions of Correlation Attacks. 
    
 In cryptanalysis one may want to distinguish non-uniform distributions from uniform. The number of zero values of $q_y$ denoted $N_q$ and the entropy of $q$ on its support denoted $H_q$ are relevant parameters. 
  For a fixed $r$ the distributions $q$ may be partitioned into  classes by equivalence, where $q_1$ and $q_2$ are equivalent if $N_{q_1}=N_{q_2}$ and $H_{q_1}=H_{q_2}$. Obviously, the number of zero values provides with a stronger distinguisher than the entropy. So we define
   an order on classes $\{q\}$ induced by the relation  
    $\{q_1\}>\{q_2\}$ which holds  if $N_{q_1}>N_{q_2}$ or if $N_{q_1}=N_{q_2}$ and $H_{q_1}<H_{q_2}$. The parameters $(N_q,H_q)$ of the largest (according to $>$) class we call $r$-dimensional non-linearity of $f$. They are  denoted $(N_f,H_f)$.  Let, for instance,  $r=n$ then $(N_f,H_f)=(2^n-v,\ln(v))$.   It is easy to see that $r$-dimensional non-linearity of $f$ is invariant under  affine change of variables in $f$. 
      
       Let $r=1$, then $Ux$ is a non-zero linear  function and 
  \begin{eqnarray}
q_0-q_1&=&\sum_xp_x\,(-1)^{Ux}=\frac{1}{v}\sum_{x}f(x)\,(-1)^{Ux}\nonumber\\
&=&\frac{1}{v}\sum_{x}\frac{1-(-1)^{f(x)}}{2}\,(-1)^{Ux}=\frac{-2^{n-1}}{v}W_U
.\nonumber
\end{eqnarray}
The numbers
$W_a=\frac{1}{2^n}\sum_{x}(-1)^{f(x)+ax}$, where $a$ are encoded by binary $n$-strings, may be called Walsh-Hadamard spectrum of $f$. Also $(-1)^{f(x)}=\sum_{a}W_a\,(-1)^{ax}$. 

 It is well known and easy to prove that  $W_a=\frac{v_a-2^{n-1}}{2^{n-1}}$, where $v_a$ is the number of $x$ such that $f(x)= ax$. 
  Minimum distance of $f$ to  affine functions (classical non-linearity of $f$)  is defined by
 $$d_f=\min_{a}(v_a,2^n-v_a)=2^{n-1}(1-\max_{a }|W_a|).$$  
 
To construct $U$, where the distribution $q_0,q_1$ has the smallest entropy (largest bias $|q_0-q_1|$), one computes  Walsh-Hadamard spectrum of $f$ and chooses $U$ such that $W_U$ is the largest in absolute value. The computation takes $n2^n$ integer additions and subtractions.
So the largest   Walsh-Hadamard spectrum value $|W_a|,a\ne 0$ is a $1$-dimensional parameter of the Boolean function $f$. For balanced Boolean functions $W_0=0$. So $1$-dimensional non-linearity parameter for a balanced Boolean function is also defined  by the classical non-linearity of  $f$.

In order to find $r$-dimensional parameters for $r\geq 2$ one can brute force all matrices $U$ (up to an  equivalence by row operations), calculate the distribution $q$, its entropy and the number of its zero values. The number of inequivalent matrices grows fast with $r$, so the calculation is infeasible even for moderate $n$.

 We consider an example.       For the Boolean function $$f(x_1,x_2,x_3,x_4,x_5)=x_1x_2x_3 +x_1x_2x_4 +x_1x_2x_5 +x_1x_4 +x_2x_5 +x_3 +x_4 +x_5$$
$r$-dimensional  parameters are shown in Table \ref{Boolean} for $r=1,2,3,4$, where $u$ is the  number of $r\times n$-matrices  $U$ up to a row equivalence, $c$ is the number of classes of equivalent distributions.   
Also the table contains a  representative $q$ of the largest (according to the order above)  class of the distributions, a matrix $U_q$ and   
the number $T_q$ of the distributions in that class, and the parameters $N_f,H_f$.  The linear transform $U$ is represented by its coordinate linear functions.        
\begin{table}[htp]
\caption{$r$-dimensional non-linearity parameters of $f$}
\begin{center}
\begin{tabular}{|c|c|c|c|c|c|c|c|}\hline

$r$& $u$&$c$& $U_q$ & $q$&  $N_f$&  $H_f$& $T_q$\\ \hline
&&&  & & && \\
$1$&31&2& $x_4+x_5$ & $\frac{3}{8},\frac{5}{8}$& 0&$0.95441$& $16$\\
&  & & && &&\\ \hline
&&&  & & && \\
$2$&155&5 &$x_3+x_5$ & $\frac{5}{16},\frac{5}{16},\frac{5}{16},\frac{1}{16}$& 0&$1.82320$& $8$\\
& & &$x_4+x_5$ & & &&  \\
 \hline
 $3$&155&7& $x_2$ & & 1&$2.65563$& $12$\\
 &&& $x_3+x_5$ & $\frac{1}{16},\frac{1}{16},\frac{3}{16},\frac{3}{16},0,\frac{1}{4},\frac{1}{8},\frac{1}{8}$ & && \\
&&& $x_4+x_5$ & & && \\
 \hline
  $4$&31&3& $x_1$ & & 6&$ 3.2500$& $1$\\
  && & $x_2$ &$\frac{1}{16},\frac{1}{16},\frac{1}{16},\frac{1}{16},0,\frac{1}{8},\frac{1}{8},0$ & && \\
 &&& $x_3+x_5$ & & && \\
&&& $x_4+x_5$ & $0,0,\frac{1}{8},\frac{1}{8},0,\frac{1}{8},0,\frac{1}{8}$ & && \\
&&&  & & && \\
 \hline
\end{tabular}
\end{center}
\label{Boolean}
\end{table}%

\section{Vectorial Boolean Functions}
A variation of the above definition may be extended to vectorial Boolean functions. Let $y=(y_1,\ldots,y_m)=f(x_1,\ldots,x_n)$ be a vectorial Boolean function in $n$ variables $x=(x_1,\ldots,x_n)$. One defines a probability distribution on $(n+m)$-binary vectors $p_{x,y}=1/2^n$ if $y=f(x)$ and $p_{x,y}=0$ otherwise. Let $U$ be an $r\times (n+m)$ binary matrix of rank $r$. That matrix defines a probability distribution $q$ on binary $r$-strings as
$q_z=\sum_{z=U(x,y)}p_{x,y}$, where the sum is computed over  $x,y$ such that $z=U(x,y)$, and $(x,y)$ is a column vector of length $n+m$.
How to  find efficiently $U$ such that the distribution $q$ is far away from the uniform? For $r=1$ the distribution is $q=(q_0,q_1)$ and the function $U(x,y)=ax+by$ is a conventional linear approximation used in Matsui's Linear Cryptanalysis of block ciphers, see \cite{Matsui}.  The best $ax+by$ is found after  applying Walsh-Hadamard transform to the distribution $p$ as
$$q_0-q_1=\sum_{x,y}p_{x,y}(-1)^{ax+by}.$$
The computation takes $(n+m)2^{n+m}$ arithmetic operations. Let $r=m=1$.  We set $b=1$, otherwise the distribution $q$ is uniform. Then
$$q_0-q_1=\sum_{x,y}p_{x,y}(-1)^{ax+y}=\frac{1}{2^n}\sum_{x}(-1)^{ax+f(x)}=W_a.$$
One takes $a$ with the largest $|W_a|$ and constructs the  distribution $q$ with the smallest entropy by using $U(x,y)=ax+y$.

 Similar to Section \ref{Boolean}, we define  $r$-dimensional non-linearity parameters $(N_f,H_f)$ for $f$.
 in case $r\geq 2$  efficient method to compute those parameters  is unknown. However for small $n$ one can  brute force all matrices $U$.   
For instance, let $n=m=4$ and $f(x_1,x_2,x_3,x_4)=(y_1,y_2,y_3,y_4)$, where 
\begin{eqnarray}
(x_1\alpha^3+x_2\alpha^2+x_3\alpha+x_4)^{-1}=(y_1\alpha^3+y_2\alpha^2+y_3\alpha+y_4)\quad \hbox{mod} \quad\alpha^4+\alpha+1
\end{eqnarray}
if $(x_1,x_2,x_3,x_4)\ne (0,0,0,0)$ and $f(0,0,0,0)=(0,0,0,0).$  $r$-dimensional non-linearity parameters for $f$ for $r=1,\ldots,7$ are in Table \ref{Vectorial}.

\begin{table}[htp]
\caption{$r$-dimensional non-linearity parameters of the vectorial $f$}
\begin{center}
\begin{tabular}{|c|c|c|c|c|c|c|c|}\hline

$r$& $u$&$c$& $U_q$ & $q$&  $N_f$&  $H_f$& $T_q$\\ \hline
&&&  & & && \\
$1$&255&3& $x_1+x_2+x_3+y_1$ & $\frac{3}{4},\frac{1}{4}$& 0&$0.8112$& $30$\\
&  & & && &&\\ \hline
&&&  & & && \\
$2$&10795&12 &$x_1+x_3+x_4+y_4,$ & $\frac{1}{2},\frac{1}{4},0,\frac{1}{4}$& 1&$1.5$& $135$\\
& & &$x_2$ & & &&  \\
 \hline
 $3$&97155&35& $x_1+y_1,$ & & 3&$2$& $15$\\
 &&& $x_2+y_1+y_2,$ & $\frac{1}{2},0,0,0,\frac{1}{8},\frac{1}{8},\frac{1}{8},\frac{1}{8}$ & && \\
&&& $x_3+y_1+y_3$ & & && \\
 \hline
  $4$&200787&49& $x_1+y_2+y_3,$ & & 10&$ 2.4056$& $3$\\
  && & $x_2+y_1+y_2+y_4,$ &$\frac{3}{8},0,\frac{1}{8},\frac{1}{8},0,0,0,\frac{1}{8}$ & && \\
 &&& $x_3+y_3+y_4,$ & & && \\
&&& $x_4+y_2$ & $0,0,0,\frac{1}{8},0,\frac{1}{8},0,0$ & && \\
&&&  & & && \\
 \hline
 $5$&97155&21& $x_1+y_2+y_3,$ &$\frac{1}{4},0,0,0,\frac{1}{16},0,\frac{1}{16},0$ & 23&$ 3$& $30$\\
  && & $x_2+y_2+y_4,$ &$0,\frac{1}{8},0,0,0,\frac{1}{16},\frac{1}{8},\frac{1}{16}$ & && \\
 &&& $x_3+y_3+y_4,$ &$0,0,0,\frac{1}{8},0,0,0,0$ & && \\
&&& $x_4+y_2$ & $0,0,0,0,0,0,0,\frac{1}{8}$ & && \\
&&& $y_1$ & & && \\
 \hline
 $6$&10795&9& $x_1+y_3,x_2,x_3,$ &$\frac{3}{16},0,0,\frac{1}{16},0,\ldots,0$ & 52&$ 3.4528$& $90$\\
&&& $x_4+y_3+y_4,y_1,y_2$ &  & && \\\hline
$7$&255&3& $x_1,x_2,x_3,$ &$\frac{1}{8},0,0,,\ldots,\frac{1}{16},0,0,0$ & 114&$ 3.75$& $15$\\
&&& $x_4+y_4,y_1,y_2,y_3$ &  & && \\
 \hline

\end{tabular}
\end{center}
\label{Vectorial}
\end{table}%
One can construct an extension to the Linear Cryptanalysis based on $r$-dimensional parameters for  $r\geq 2$. 
\section{Optimal Boolean Functions}\label{optimal}
Let $y=f(x_1,\ldots,x_n)$ be a vectorial Boolean function with $m$ coordinates and in $n$ variables and let $1\leq r\leq n$. 
One can split Boolean functions with the same $n,m$ into classes of equivalence and define an order on the equivalence classes.

 Let $f'$ be another Boolean function with the same parameters $n,m$. One says $f,f'$ are equivalent (belong to the same class denoted $\{f\}$) if $(N_f,H_f)=(N_{f'},H_{f'})$. One now defines an order on the classes by 
$\{f\}<\{f'\}$ if $N_{f}<N_{f'}$ or if $N_{f}=N_{f'}$, then $H_{f}>H_{f'}$. Boolean functions from the smallest (according to $<$) class are called optimal. 

 Then it is easy to show that for  $r=1,m=1$ and even $n$  optimal Boolean functions     are exactly  Boolean bent functions introduced by Rothaus in \cite{R76}. For $m\geq 1$ and  even $n\geq 2m$ optimal Boolean functions are perfect nonlinear according to \cite{N91} and vice versa. By a computer search we find that the property  holds for $r=2, m=1, n=4$. For larger $n,m$ and $r\geq 2$ this is an open problem. 

\bibliographystyle{alpha}

\end{document}